\documentclass[aps,prl,twocolumn,showpacs,preprintnumbers,amsmath,amssymb]{revtex4}

\usepackage{graphicx}
\usepackage{dcolumn}
\usepackage{bm}
\begin{document}

\title{Andreev Reflection in Heavy-Fermion Superconductors and Order Parameter Symmetry in CeCoIn$_5$}

\author{W. K. Park,$^1$ J. L. Sarrao,$^2$ J. D. Thompson,$^2$ and L. H. Greene$^1$}
\affiliation{$^1$Department of Physics and Frederick Seitz Materials Research Laboratory, University of Illinois at Urbana-Champaign, Urbana, Illinois 61801, USA \\
$^2$Los Alamos National Laboratory, Los Alamos, New Mexico 87545, USA}

\date{\today}

\begin{abstract}
Differential conductance spectra are obtained from nanoscale junctions on the heavy-fermion superconductor CeCoIn$_5$ along three major crystallographic orientations. Consistency and reproducibility of characteristic features among the junctions ensure their spectroscopic nature. All junctions show a similar conductance asymmetry and Andreev reflection-like conductance with reduced signal ($\sim 10\%-13$\%), both commonly observed in heavy-fermion superconductor junctions. Analysis using the extended Blonder-Tinkham-Klapwijk model indicates that our data provide the first spectroscopic evidence for $d_{x^2-y^2}$ symmetry. To quantify our conductance spectra, we propose a model by considering the general phenomenology in heavy fermions, the two-fluid behavior, and an energy-dependent density of states. Our model fits to the experimental data remarkably well and should invigorate further investigations.
\end{abstract}


\maketitle

The 1-1-5 family of heavy fermion (HF) compounds, Ce$T$In$_5$ ($T$ = Co, Rh, Ir), has attracted great interest because of their novel and rich physical phenomena \cite{thompson03}. Various thermodynamic and transport measurements indicate that the superconducting order parameter (OP) in CeCoIn$_5$ is $d$-wave \cite{thompson03}, but precise locations of the line nodes over the Fermi surface remain controversial \cite{izawa01,aoki04}. Not only is the interpretation of these experiments complex \cite{vorontsov06}, but also they intrinsically cannot provide phase information of the OP. In this Letter, we report differential conductance data on CeCoIn$_5$ as a function of temperature, magnetic field, and crystallographic orientation. Our results show the first spectroscopic evidence for $d_{x^2-y^2}$ symmetry. We further present a model, which, for the first time, quantifies the Andreev signal and conductance asymmetry observed in normal-metal/heavy-fermion superconductor (N/HFS) junctions \cite{naidyuk98,park05,goll06,park07physc}.

Andreev reflection (AR), the scattering of a quasiparticle off an attractive pair potential, occurs at an N/S interface as the retro-reflection of an electron as a hole \cite{andreev64}. If the N and S are in good electrical contact and their Fermi velocities are well matched, the conductance is twice the normal state value within the superconducting energy gap, $\Delta$, and rapidly returns to the normal state value outside $\Delta$, directly providing energy gap information. The Blonder-Tinkham-Klapwijk (BTK) theory \cite{btk} describes the conductance data in practical N/S junctions remarkably well using a dimensionless barrier strength parameter, $Z_\textrm{eff} = [Z_0^2 + (1-r)^2/4r]^{1/2}$, where $Z_0$ is due to a physical potential barrier and $r$ is the ratio of Fermi velocities  \cite{btk}. As $Z_\textrm{eff}$ increases from zero, the junction moves from the AR to the tunneling regime. In HF systems, the electronic mass is highly enhanced (by $\sim 10-10^3$) with a correspondingly reduced Fermi velocity. According to the above formula, a N/HFS junction is inherently in the tunneling regime and AR cannot occur. However, AR is frequently observed in N/HFS junctions (Ref. \cite{naidyuk98} and references therein). Deutscher and Nozi{\'e}res addressed this discrepancy by assuming the boundary conditions are not affected by mass enhancement \cite{deutscher94}.

\begin{figure}[t]
\includegraphics[scale=0.87]{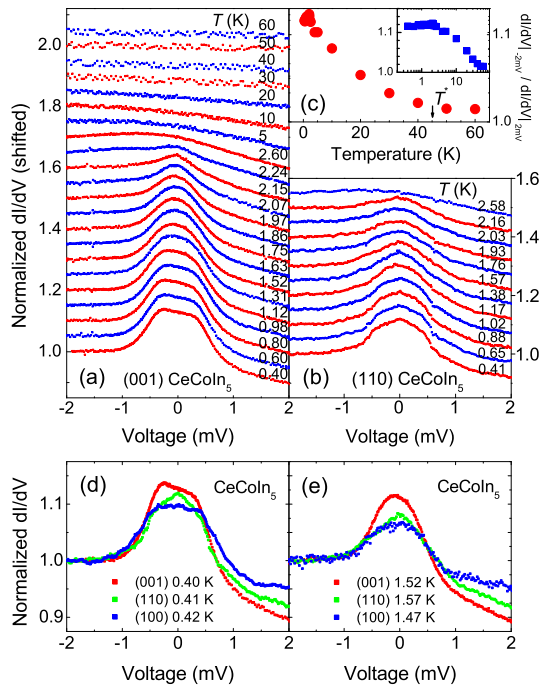}
\caption{\label{fig:fig1} (color online). Normalized conductance spectra of CeCoIn$_5$/Au junctions (a) along (001) (after Ref. \cite{park05}) and (b) along (110) orientations. Data are shifted vertically for clarity. Note the temperature evolution of the background conductance, whose asymmetry is quantified in (c) by the ratio between conductance values at --2 mV and at +2 mV in (a); the inset is a semi-logarithmic plot ($T^*$ is the HF coherence temperature). Junctions along three orientations are compared in (d) at $\sim$ 400 mK and in (e) at $\sim$ 1.5 K.}
\end{figure}

Three crystallographic surfaces of high quality CeCoIn$_5$ single crystals are prepared: the (001) face of as-grown crystals, and the (110) and (100) surfaces by embedding into epoxy and polishing. X-ray diffraction analysis confirms their crystallographic orientations \cite{park07physc}. A light HCl etch exposes fresh surfaces prior to junction formation. The average surface roughness, which ranges 1 $-$ 2 nm over $\sim$ 1 $\mu$m$^2$ area, is much smaller than the junction size, estimated below. Electrochemically polished metal tips, Au (primarily), Al, or Nb, are used as counter-electrodes. Nanoscale junctions are formed by electromechanical adjustment of the tip-sample distance in our Cantilever-Andreev-Tunneling rig \cite{park06rsi} that is run in a $^3$He fridge. Differential conductance (dI/dV) data are taken using a standard four probe lock-in technique over wide ranges of temperature (400 mK $-$ 60 K) and magnetic field (0 $-$ 9 Tesla). High-bias junction resistances typically range 1 $-$ 5 $\Omega$, which correspond to junction sizes of 20 $-$ 50 nm, estimated using Wexler's formula \cite{wexler66}, indicating that junctions are in the ballistic (Sharvin) limit; the extreme cleanness of this compound makes it readily accessible (the electronic mean free path ranges several micrometers at low temperature \cite{kasahara05}). However, we stress this is not a sufficient condition for AR spectroscopy (see Ref. \cite{park06prl} and references therein). Reproducibility and consistency of the conductance spectra along different crystallographic orientations are crucial to ensure their intrinsic and spectroscopic nature.

Normalized conductance spectra for the (001) and (110) junctions are displayed in Figs. 1(a) and 1(b), respectively. At high temperatures, the conductance curves of the (001) junction are symmetric and flat; characteristic of simple metallic junctions. As the temperature is reduced, they become asymmetric and curved. This conductance asymmetry begins at the HF coherence temperature $T^*$ ($\sim$ 45 K) \cite{nakatsuji04} and increases with decreasing temperature down to $T_\textrm{c}$ (2.3 K), below which it remains constant. The same behavior is observed in the (110) junction and the data near and below $T_\textrm{c}$ are shown in Fig. 1(b). A plot of the ratio of the conductance values at --2 mV and +2 mV quantifies this asymmetry (Fig. 1(c)). According to the two-fluid model proposed by Nakatsuji, Pines and Fisk \cite{nakatsuji04}, the spectral weight for the emerging HF liquid grows below $T^*$ and saturates below $T_\textrm{c}$ \cite{nakatsuji04}, and our conductance data track this behavior.

The conductance near zero-bias begins to be enhanced as $T_\textrm{c}$ is crossed and increases with decreasing temperature, indicating its origin is AR. Conductance data at two temperatures are compared for all three orientations in Figs. 1(d) and 1(e). Three consistent and reproducible characteristics are observed at low temperature, indicating we are sampling $intrinsic$ $spectroscopic$ properties. First, all spectra are asymmetric with the positive-bias side (electrons flowing into CeCoIn$_5$) always lower than the negative-bias branch. We have seen similar conductance asymmetry from more than two hundred junctions on pure and Cd-doped CeCoIn$_5$ along all three directions. This is in strong contrast with the symmetric conductance data we obtained from junctions on non-HFS such as Nb, MgB$_2$ \cite{park06rsi}, and LuNi$_2$B$_2$C. All these observations strongly indicate that the conductance asymmetry arises from intrinsic properties in CeCoIn$_5$. Second, the conductance enhancement occurs over similar voltage ranges, $\sim \pm(1 - 1.5)$ mV. Third, the normalized zero-bias conductance (ZBC) ranges 1.10 $-$ 1.13, showing that our observed Andreev signal is much smaller than the theoretical prediction of 100\% \cite{btk}. We reported \cite{park05} that it is too small to fully account for the conductance spectra using the existing BTK models even considering the mismatch in Fermi surface parameters, nonzero $Z_\textrm{eff}$, and large quasiparticle lifetime broadening factor ($\Gamma$). Our model proposed below enables us to quantify it successfully and elucidates properties of the HFS state.

\begin{figure}[t]
\includegraphics[scale=0.87]{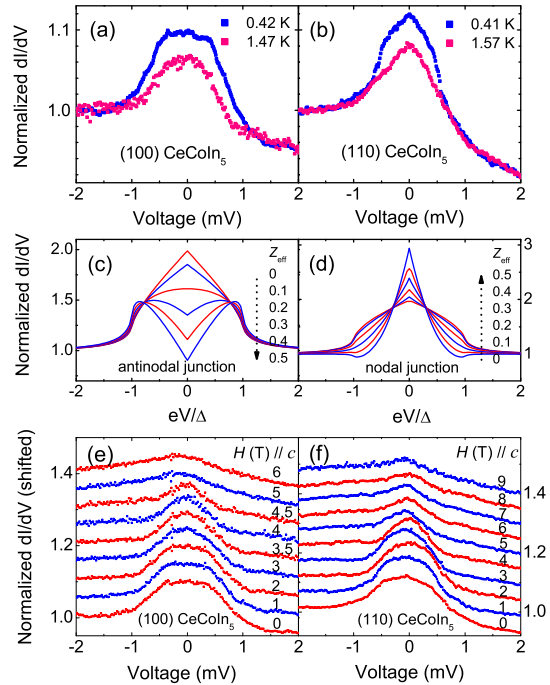}
\caption{\label{fig:fig2} (color online). Comparison of conductance data: (a) the (100) and (b) the (110) junctions. (c) and (d) Calculated conductance curves using the $d$-wave BTK model ($\Gamma$ = 0 and $T=0$) for antinodal and nodal junctions, respectively. (e) and (f) Magnetic field dependence for the (100) junction at 400 mK and the (110) junction at 420 mK, respectively.}
\end{figure}

Conductance spectra for in-plane junctions are plotted in Figs. 2(a) and 2(b). While both spectra exhibit similar background asymmetry, differences in the sub-gap region are striking. The (100) data appear rather flat, similar to the (001) junction, whereas the (110) data are cusp-like. This shape difference persists even to higher temperature despite an enhanced thermal population effect, indicating it is intrinsic. We compare these data with calculated conductance curves using the $d$-wave BTK model \cite{kashiwaya96}, as shown in Figs. 2(c) and 2(d) for antinodal and nodal junctions, respectively. Both curves are identical at $Z_\textrm{eff} = 0$ but quickly evolve in dramatically different manners with increasing $Z_\textrm{eff}$.  For an antinodal junction, the ZBC is gradually suppressed and a double-peak structure develops for $Z_\textrm{eff}\sim 0.3$. For a nodal junction, the ZBC increases and the sub-gap conductance narrows into a sharp peak. This is the signature of Andreev bound states (ABS) which arise directly from the sign change of the OP around the Fermi surface \cite{hu94}. We stress the flat conductance shape observed in the (100) junction can only occur for an antinodal junction with $Z_\textrm{eff}=0.25-0.30$ but cannot occur in a nodal junction at any $Z_\textrm{eff}$ value. Meanwhile, the cusp-like feature in the (110) junction cannot occur in an antinodal junction unless $Z_\textrm{eff}$ is small enough ($\sim$ 0.1), an unlikely condition in N/HFS junctions; it can only be explained by a sign change of the OP, ruling out anisotropic $s$-wave. We therefore assign the (100) and (110) orientations as the antinodal and nodal directions, respectively, providing evidences for $d_{x^2-y^2}$-wave symmetry and resolving the controversy on the locations of the line nodes \cite{izawa01,aoki04,vorontsov06}. Note this is a spectroscopic measurement of the superconducting OP symmetry in that it can detect its sign change, in contrast with other measurements that probe only the gap anisotropy, including heat transport and NMR. The ABS-originated ZBC peaks are reported to split spontaneously and/or under applied magnetic field in high-$T_\textrm{c}$ cuprate tunnel junctions \cite{covington97}. One of the widely adopted explanations is the Doppler shift of ABS. We test it by applying a magnetic field perpendicular to the $ab$ plane, a configuration for a maximal shift, if any. As shown in Figs. 2(e) and 2(f), no splitting but only gradual suppression of the Andreev signal is observed in both junctions. We have observed that the $Z_\textrm{eff}$ values do not change significantly by using different tips (e.g., see Fig. 4(b)), samples, and contact pressures. While further studies are necessary, diminution of the Doppler effect due to large junction transparency, small tunneling cone, and atomic-scale disorder, has been suggested to explain similar behaviors in some cuprate junctions \cite{ABSsplitting}.

\begin{figure}[t]
\includegraphics[scale=0.85]{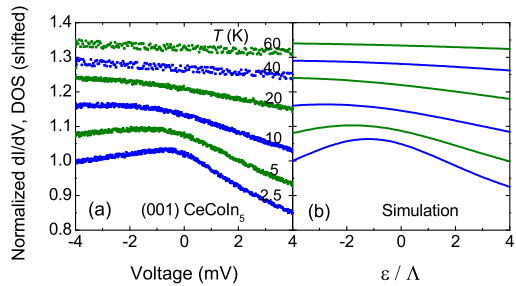}
\caption{\label{fig:fig3} (color online). (a) Normal state conductance spectra of the (001) junction, whose features are reproduced qualitatively by the simulated Lorentzian DOS curves in (b).}
\end{figure}

Both asymmetric conductance and a reduced Andreev signal have been commonly reported by others in CeCoIn$_5$  \cite{goll06} and other HFS junctions \cite{naidyuk98}. The temperature dependence of the conductance asymmetry (Fig. 1(c)) qualitatively follows that of the spectral weight of the coherent HF liquid in the two-fluid model \cite{nakatsuji04} proposed to be general to HFs \cite{curro04yang07}. These observations strongly suggest a universal mechanism to explain the charge transport at HF interfaces. We attribute it to the emergent HF liquid: with decreasing temperature, the logarithmic increase of the electronic specific heat coefficient \cite{nakatsuji04} signals the increase of the electronic mass and, equivalently, the electronic density of states (DOS) at the Fermi level. The electrical conductance of a clean metallic junction, where one electrode is a simple metal, is given by: $\frac{dI}{dV}(V) \propto \iint vD(\epsilon)\frac{\partial f(\epsilon - eV)}{\partial (eV)} d\epsilon d\Omega$, where $v$ is the velocity, $D$($\epsilon$) the DOS of the counter-electrode, $f$ the Fermi-Dirac distribution function, and d$\Omega$ the differential solid angle \cite{naidyuk98}. For a simple metal, the DOS is constant around the Fermi level and thus divides out in the normalized conductance giving flat and symmetric shape. For a HF metal, the energy-dependent DOS is reflected in the conductance data \cite{nowack92}; an asymmetric DOS yields asymmetric conductance. In Fig. 3(a), our normal state conductance data exhibit an increasing asymmetry with decreasing temperature while the conductance peak sharpens and shifts towards the Fermi level. We have investigated several proposed models in reproducing this non-trivial temperature dependence, including large Seebeck effect and non-Fermi liquid behavior in HFs. We find that only our model, in which we assume a peak in the DOS below the Fermi level, reproduces the observed experimental features (Fig. 3(b)) {\cite{park-prb}}.

\begin{figure}[t]
\includegraphics[scale=0.87]{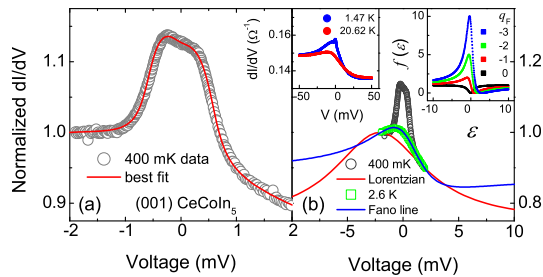}
\caption{\label{fig:fig4} (color online). (a) Best fit to the data using our modified BTK model. (b) Comparison of the data with calculated DOS curves. Right inset: Fano lines simulated by $f(\epsilon) = (q_F+\epsilon)^2 / (1+\epsilon^2)$, where $q_\textrm{F}$ is the Fano factor and $\epsilon$ normalized energy. Left inset:  representative conductance data of Fano line shape taken with Al tips.} 
\end{figure}

Measurements of de Haas-van Alphen effect in CeCoIn$_5$ show that heavy and light fermions reside on disparate Fermi surfaces \cite{shishido02}. Tanatar {\it et al}. reported \cite{tanatar05} that the light electrons remain uncondensed below $T_\textrm{c}$. The junction conductance would therefore be a measure of both normal and superconducting channels. We modify the BTK model such that the total conductance is given by the sum of two parallel conductance channels: $\frac{dI}{dV} (V) = \omega_h \frac{dI}{dV} \big|_{h} (V) + (1-\omega_h) \frac{dI}{dV} \big|_{l} (V)$, where the first term is the usual BTK conductance \cite{kashiwaya96} due to the superconducting heavy electrons, the second a constant conductance due to the uncondensed light electrons, and $\omega_h$ the weighting factor related to the HF spectral weight \cite{nakatsuji04}. The DOS for the HFs, used in the BTK conductance kernel, is modeled as a Lorentzian centered at $\epsilon_0$: $D(\epsilon) = D_0 \big[1 + \eta\frac{\Lambda^2}{\Lambda^2+(\epsilon-\epsilon_0)^2} \big]$, where $D_0$ is a constant, and $\eta$ and $\Lambda$ are the peak height and half width, respectively.

Figure 4(a) shows the best-fit curve for the (001) data taken at 400 mK with $\omega_h$ = 0.51, $\Delta$ = 600 $\mu$eV, $\Gamma$ = 95 $\mu$eV, $Z_\textrm{eff}$ = 0.28, $\eta$ = 1, $\Lambda$ = 5 meV and $\epsilon_0$ = --2.1 meV. The physical origin of the peak in the DOS below the Fermi level remains an open question, particularly with direct measurements such as photoemission or tunneling still lacking. Compared to our previous fit \cite{park05} using the single channel BTK model \cite{kashiwaya96}, the quality of the fit is remarkable; it nicely reproduces both features of a reduced AR signal and the conductance asymmetry. The obtained energy gap gives 2$\Delta/k_\textrm{B}T_\textrm{c} = 6.05$, suggesting strong coupling in agreement with literature. We find the fit using our proposed model is particularly sensitive to $\omega_h$; for smaller values it becomes much poorer; for larger values it requires smaller $\Delta$ and larger $\Gamma$ values, causing the same problem of an unphysical temperature dependence of $\Gamma$ as in our previous analysis \cite{park05}. Thus, both superconducting and normal conductance channels are necessary. Our model provides a natural explanation for the unreduced Andreev signal in our CeCoIn$_5$/Nb junctions at temperatures between the two $T_\textrm{c}$'s \cite{park07}; here both heavy and light electrons in CeCoIn$_5$ participate in AR. In the high temperature region, the fitted curve deviates substantially from the data and we find that a Fano \cite{fano61} line shape as a background gives a much better fit, as illustrated in Fig. 4(b). The Fano line as a function of the Fano factor, $q_\textrm{F}$, is shown in the right inset of Fig. 4(b). The line shape for $q_\textrm{F} \sim$ --2 is similar to those observed in various junctions by Goll {\it et al}. \cite{goll06} and by us over wide voltage ranges, as exemplified in the left inset of Fig. 4(b). Since the physical origin of the Fano-like background needs to be clarified further, we defer quantitative analyses of the complete conductance spectra \cite{park-prb}. As seen, our model correctly captures the underlying physical phenomena. This implies that microscopic details such as boundary conditions do not play a key role in reducing an Andreev signal. Since the two-fluid behavior is proposed to be universal to HFs \cite{curro04yang07}, our model should also be generally applicable.

W.K.P. and L.H.G. are grateful to A. J. Leggett, D. Pines, and V. Lukic for fruitful discussions and X. Lu for experimental help. Work at the UIUC was supported by the U.S. DoE, Grant DEFG02-91ER45439, through the FSMRL and the CMM. Work at LANL was performed under the auspices of the U.S. DoE Office of Science.

\end{document}